\RequirePackage[2020-02-02]{latexrelease}
\documentclass[twocolumn,preprintnumbers,amsmath,amssymb]{revtex4}
\usepackage{graphicx}
\usepackage{dcolumn}
\usepackage{bm}
\usepackage{color}
\usepackage{physics,dsfont}
\usepackage{subfigure}
\usepackage{ulem}

\begin{document}
	
	\title{Shortcut to synchronization in classical and quantum systems}
	
	\author{Fran\c{c}ois Impens$^1$ and David Gu\'ery-Odelin$^2$}
\affiliation{$^1$ Instituto de F\'{i}sica, Universidade Federal do Rio de Janeiro,  Rio de Janeiro, RJ 21941-972, Brazil
\\
$^2$ Laboratoire Collisions, Agr\'egats, R\'eactivit\'e, FeRMI, Universit\'e de Toulouse, CNRS, UPS, France}

	\begin{abstract}
		Synchronization is a major concept in nonlinear physics. In a large number of systems, it is observed at long times for a sinusoidal excitation. In this paper, we design a transiently non-sinusoidal driving to reach the synchronization regime more quickly. We exemplify an inverse engineering method to solve this issue on the classical Van der Pol oscillator. This approach cannot be directly transposed to the quantum case as the system is no longer point-like in phase space. We explain how to adapt  our method by an iterative procedure to account for the finite-size quantum distribution in phase space. We show that the resulting driving yields a density matrix close to the synchronized one according to the trace distance. Our method provides an example of fast control of a nonlinear quantum system, and raises the question of the quantum speed limit concept in the presence of nonlinearities.
	\end{abstract}

\maketitle

The synchronisation of dynamical systems is a broad, multidisciplinar field with important applications in basic science and technology, including among other areas  biology, physics, chemistry, and engineering~\cite{PikovskyBook}. Since the pioneering work by Van der Pol on self-sustained oscillators~\cite{vanderPol}, the synchronization between coupled systems has been discussed extensively, in particular in the context of the Kuramoto model~\cite{RMPKuramoto}. In the last decades, synchronization has been transposed to the quantum realm~\cite{QuantumSynchro1,QuantumSynchro2,QuantumSynchro3,QuantumSynchro4,QuantumSynchro5,QuantumSynchro6,QuantumSynchro7,Lee13,QuantumSynchro9,Lee14,QuantumSynchro11,QuantumSynchro12,QuantumSynchro13,QuantumSynchro14,QuantumSynchro15,QuantumSynchroGiovanetti15,QuantumSynchro16,QuantumSynchro17,QuantumSynchro18,QuantumSynchro19,QuantumSynchro20,QuantumSynchroMok20,QuantumSynchroJaseem20,QuantumSynchro21,QuantumSynchro22noise,QuantumSynchro22exp,QuantumSynchro22Topo,QuantumSynchro22clas}. Quantum synchronization in ion traps~\cite{Lee13} has been implemented experimentally~\cite{QuantumSynchro22Topo} thanks to 
 controllable gain and losses. However, fundamental features of quantum mechanics, such as the quantum noise arising from Heisenberg's uncertainty principle, can induce significant qualitative differences with respect to their classical counterpart.

 The synchronization of a system driven by an external force is usually formulated as follows: a sinusoidal driving is suddenly applied and one investigates the asymptotic behavior of the driven system in the long-time limit. The driving frequency is  detuned with respect to the natural frequency of the system. Synchronization is achieved when the driven system locks on the driving frequency. For a given detuning, the synchronization process requires a sufficient large driving amplitude, a feature often pictured as Arnold's tongues~\cite{PikovskyBook}. 
 Synchronization is therefore considered essentially an asymptotic phenomenon, and most studies have focused on determining the domain of parameters associated with the onset of synchronization without explicitly discussing the pace at which it takes place.
  
The issue of the synchronization time is relevant for several practical purposes. Indeed, an acceleration of quantum protocols increases their quantum fidelity by reducing the detrimental influence of decoherence. The acceleration of quantum state transformations is at the heart of the fields of optimal quantum control  \cite{oct} and shortcut to adiabaticity (STA)  \cite{STARMP}. However,  most STA techniques exploits the linearity of the Schr\"{o}dinger equation,~as illustrated by numerous applications to simple quantum systems~\cite{Impens17,Impens19}. %as illustrated by several applications of STAs to simple quantum systems~
 The extension to nonlinear quantum systems is at the incipient stage \cite{Muga09,Zhang15,Liu20,Zhu20,Huang20,Kong20} and raises the question of the influence of non linearities on quantum speed limit~\cite{NonlinearDeffner22}. In this paper, we propose to accelerate classical and quantum synchronization, an inherently non-linear and non-perturbative phenomenon~\cite{PikovskyBook}. Speeding-up quantum synchronization translates into the control of an extended quantum object under a nonlinear dynamics.  

 This paper is organized as follows. We first discuss the acceleration of synchronization in a classical Van der Pol oscillator. We then consider the analogous quantum system, whose density matrix follows a Master equation with a pump and a gain reproducing the Van der Pol oscillator in the classical limit~\cite{Lee13}.  We point out a key difference  between classical and quantum synchronization. While classical synchronization is obtained by driving the system position to a single point of the limit cycle, quantum synchronization requires that the full system density matrix matches the stationary solution of the Master equation. Furthermore, quantum synchronization is a nonlinear process that typically produces highly non-classical  states. Thus, building a shortcut to a perfect quantum synchronization, i.e. designing time-dependent control parameters that bring the system's density matrix to its target in a finite time is a very challenging (and possibly intractable) task. To circumvent this issue, we outline a simpler strategy for approaching the target density matrix based on the mean position and lowest-order moments. We show that shortcuts  based on this method provide a strong acceleration towards quantum synchronisation. 

\textit{Accelerated synchronization of a classical Van der Pol oscillator} - First consider a Van der Pol oscillator~\cite{vanderPol} driven by an external sinusoidal force. In the weakly nonlinear regime, the system dynamics boils down to a nonlinear first-order differential equation for a complex-valued function $\alpha(t))$~\cite{PikovskyBook}:
 \begin{equation}
 	\label{eq:classicalvanderPol}
 	\dot{\alpha} = - i \omega_0 \alpha+  \alpha (\kappa_1 -2 \kappa_2 |\alpha|^2) - i \frac { \epsilon(t)} {2}
 \end{equation}
The system follows a 2D trajectory $(x(t)={\rm Re}[\alpha(t)]$, $y(t)={\rm Im}[\alpha(t)])$  in the complex plane.
The driving only acts on the $y$ coordinate, and in the usual formulation of the synchronization problem, one considers a sinusoidal driving $\epsilon_0(t) = \epsilon_0 \cos(\omega t ) $. With such a driving, the Van der Pol oscillator converges to a limit cycle asymptotically. It therefore takes an arbritrarily long time to approach the limit cycle within an arbitrary close neighborhood. We propose to go beyond this approach and to study the behavior of the system for a more general class $\epsilon(t)$ of driving functions involving transiently a non-sinusoidal profile.  We present below a procedure to speed up the synchronization for a given coupling strength $\epsilon_0$ and frequency $\omega$ (compatible with the onset of synchronisation) and for a given initial system position $(x_0,y_0)$.

 In the following, we build up a piece-wise driving with a sinusoidal form of given amplitude and frequency $\epsilon(t)= \epsilon_0 \cos (\omega t  + \varphi )$ for $ t > \tau$, and explain how to design the driving $\epsilon_{\rm short}(t)$, referred to as the shortcut driving,  in the time interval $0 \leq t \leq \tau$, to reach the limit cycle. With our method, the convergence of the system trajectory to the limit cycle is no longer asymptotic - it occurs over a short and finite time scale, which can be significantly shorter than an oscillation cycle. With the properly designed $\epsilon_{\rm short}(t)$ and a suitable phase $\varphi$, the system evolves on the limit cycle for $t>\tau$.
 
First, we consider the system trajectory $(x_0(t),y_0(t))$ under the sinusoidal drive $\epsilon_0(t)= \epsilon_0  \Theta(t) \cos (\omega t)$ 
 and identify a point of the limit cycle, where $\Theta$ is the step Heaviside function. In practice, one can solve numerically  Eq.~\eqref{eq:classicalvanderPol} and choose a late time $t_{\infty} \gg 2 \pi/ \omega$ for which the system position $(x_0(t_{\infty}),y_0(t_{\infty}))=(x_{\infty},y_{\infty})$ is already extremely close to the limit cycle. This position defines a branching point for the shortcut trajectory. This choice is by no means unique - one could consider, in principle, any point of the limit cycle, and we shall see below that the ideal branching point depends indeed on the initial coordinates $(x_0=x_0(0),y_0=y_0(0))$. The driving $\epsilon(t)$ provides a precise control of the $y$ coordinate of the system. One can thus design an arbitrary trajectory $y_{\rm short}(t)$ between $y_0$ and $y_{\infty}$ in the interval $[0,\tau]$, i.e. a trajectory that fulfills the boundary conditions $y_{\rm short}(0)=y_0$ and $y_{\rm short}(\tau)=y_{\infty}$. From Eq.\eqref{eq:classicalvanderPol}, we deduce that a self-consistent solution requires to solve for $x(t)$ a nonlinear differential equation where the chosen path $y_{\rm short}(t)$ plays the role of a source term. By plugging in Eq.~\eqref{eq:classicalvanderPol} the corresponding solution, $\alpha_{\rm short}(t)=x_{\rm short}(t)+i y_{\rm short}(t)$, we determine self-consistently the required driving $\epsilon_{\rm short}(t)$~\cite{note1}. Finally, we set the phase $\varphi= \omega (t_{\infty}-\tau)$, so that $\epsilon(\tau)=\epsilon_0(t_{\infty})$. Then, the shortcut-driven system reaches at $t=\tau$ the position occupied by the sinusoidally driven system at $t=t_{\infty}$, and is subject to an identical driving at later times. By uniqueness of the solution, the system driven by the design driving $\epsilon_{\rm short}(t)$ subsequently evolves (for $t \geq \tau$) in a neighborhood extremely close to the limit cycle.

 There is, however, a significant difference between the coordinates paths $x_{\rm short}(t)$ (``slave'' coordinate) and $y_{\rm short}(t)$ (``pilot'' coordinate). While the path $y_{\rm short}(t)$ ends up by construction at the target $y_{\infty}$ at  time $t=\tau$, nothing guarantees that $x_{\rm short}(t)$ reaches the target $x_{\infty}$ at the same time. The position $x_{\rm short}(\tau)$ depends in a non-trivial way on the trajectory $y_{\rm short}(t)$ during the time interval $[0,\tau]$.  A possible way to circumvent this problem is to consider a family of possible trajectories $y_{\rm short, \gamma}(t)$ depending on a continuous parameter $\gamma$, and to select a specific parameter value $\gamma_0$ that gives $x_{\rm short,\gamma_0}(\tau)=x_{\infty}$. For a given duration $\tau$, each initial conditions $(x_0,y_0)$ admits a set of possible branching points $(x_{\infty},y_{\infty})$ on the limit cycle that can be connected by the above procedure. This set of admissible branching points covers a narrower part of the limit cycle as the shortcut duration $\tau$ is reduced.  In this regard, we note that the ``pilot'' coordinate $y_{\rm short, \gamma}(t)$ can move arbitrarily fast as long as we use a driving, $\epsilon_{\rm short}(t)$, of sufficient magnitude. Moreover, the $x$ coordinate is bounded by the duration $\tau$: $|x_{\rm short, \gamma}(\tau)-x_0| \leq  |\alpha|_{\rm max}(\omega_0+ \kappa_1+  2 \kappa_2 |\alpha|_{\rm max}^{2} ) \tau$ with $|\alpha|_{\rm max}= {\rm max} \{ |\alpha(t)| | t \in [0,\tau] \} $. If the $y$ coordinate occurs on a finite scale, that is if $|\alpha|_{\rm max} \simeq 1$, then the maximum displacement along $x$ is on the order $O(\tau).$ Thus, as the shortcut time $\tau$ is reduced, the admissible branching points $(x_{\infty},y_{\infty})$ are nearly at the ``vertical'' of the starting point $(x_0,y_0)$.
 
	A possible choice is to use a set of polynomial trajectories $y_{\rm short, \gamma}(t)=  P(t/\tau)$. Beyond the boundary conditions $y_{\rm short, \gamma}(0)=y_0$ and $y_{\rm short, \gamma}(\tau)=y_{\infty},$ we additionally impose $y_{\rm short, \gamma}'(\tau)=y_0'(t_{\infty})$ to enforce the continuity of the driving $\epsilon(t)$ at time $\tau$. The following family of polynomials  $P_{\gamma}(u)= y_{\infty}+ y_0'(t_{\infty}) \tau (u-1)+(y_0-y_{\infty}+ y_0'(t_{\infty}) \tau )(u-1)^2+\gamma u(u-1)^2$ obeys those boundary conditions independently of the $\gamma$ parameter value. This latter parameter is fixed to the value $\gamma_0$ which fulfills the condition $x_{\rm short,\gamma_0}(\tau)=x_{\infty}$.
	
	 Figures~\ref{fig:classicalsynchro} illustrate our method on a concrete example (See Appendix A for details). We use the time-scale $T_0= 2 \pi / \omega_0$ associated to the free-oscillator frequency $\omega_0$ to recast the equations in a dimensionless form. 
	 We consider $\kappa_{1}=1/T_0$, $\kappa_{2}=0.5 /T_0$, a driving amplitude $\epsilon_0=1.5 /T_0$ and a driving frequency $\omega=1.05 \times \omega_0.$ In our numerical example, we have taken an arbitrary value for the distant time $t_{\infty}=50.125 \times T_0$, which defines a possible branching point close to the $(Oy)$ axis. A fine-tuning of the constant phase $\varphi$ associated to the sinusoidal driving at times $t \geq \tau$ then guarantees that the trajectory follows on the cycle. Figure ~\ref{fig:classicalsynchro}(a) sketches the trajectory under a sinusoidal driving, showing that the limit cycle is approached gradually after a large number of cycles, and Fig.~~\ref{fig:classicalsynchro}(b) plots the trajectory under the shortcut driving. By construction, the selected branching point $(x_{\infty}, y_{\infty})$ is reached at time $\tau=T_0/4$, and the subsequent evolution occurs on the limit cycle. Figure ~\ref{fig:classicalsynchro}(c) shows the associated driving profile $\epsilon(t)$ - the shortcut part has a significantly larger amplitude. Finally, Fig.~\ref{fig:classicalsynchro}(d) represents the phase difference $\Delta \phi= \phi(t) - (\omega t + \varphi) $ between the van-der-Pol oscillator phase defined as $\phi(t)= {\rm Arctan} \left( \frac {y(t)} {x(t) } \right)$, and the driving phase $\omega t + \varphi.$ It reveals that phase locking is achieved with the engineered driving as fast as $t \simeq \tau$ up to small residual oscillations of frequency $2 \omega$. These residual oscillations also persist asymptotically when the usual sudden sinusoidal driving is applied, as synchronization occurs in the regime of slow phase dynamics~\cite{PikovskyBook}.

\begin{figure}
	\includegraphics[width=8.5 cm]{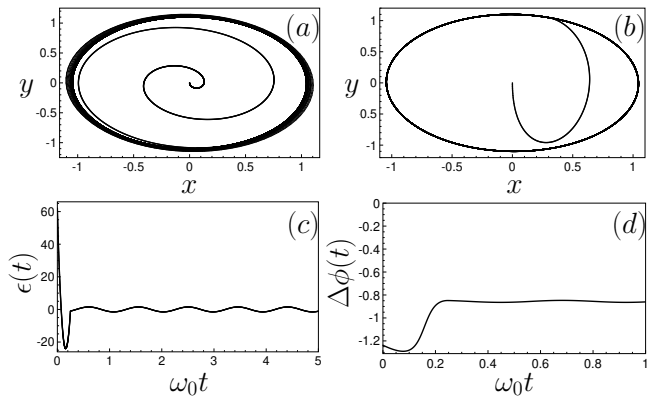}  
		\caption{Speed-up of classical synchronization. (a): System trajectory in the $(x,y)$ plane under a sudden sinusoidal driving. (b) System trajectory under a shortcut+sinusoidal driving $\epsilon(t)$. (c) Profile of the driving amplitude $\epsilon(t)$. (d) Phase difference between the shortcut-driven system and the sinusoidal drive phase $\Delta \phi=\phi(t)- \omega t - \varphi.$ ([$\pi$]).
		Parameters used:  $\omega=1.05 \times \omega_0$, $\tilde{\epsilon}_0=1.5$, $\tilde{\kappa}_1=1,$ $\tilde{\kappa}_2=0.5$ and times $\tau=T_0/4$ and $t_{\infty}=50.125 T_0$, where $T_0$ is the free oscillator period.}
	 \label{fig:classicalsynchro}
\end{figure}

\textit{Accelerated synchronization of a quantum Van der Pol oscillator}. The quantum Van der Pol oscillator was introduced by Lee et al.~\cite{Lee13}, and has become the paradigmatic model for studying synchronization in the quantum context~\cite{QuantumSynchro9,Lee14,QuantumSynchro12,QuantumSynchroGiovanetti15,QuantumSynchro16,QuantumSynchroMok20,QuantumSynchroJaseem20,vanderPolDutta19}. The quantum Van der Pol oscillator, in its original formulation, has not yet been implemented experimentally, but an important step has been taken in this direction in trapped ion physics~\cite{QuantumSynchro22exp}. Mathematically, the quantum Van der Pol oscillator is obtained by quantizing the classical Van der Pol equations. Physically, this model describes a single-mode harmonic oscillator  subject to 1- and 2-photon dampings.  The balance between these two dissipative processes and the external driving determines the steady state. The corresponding Hamiltonian reads $\tilde{H}(t) = \omega_0 a^{\dagger} a + \frac 1 2 [(\epsilon_1(t)+i \epsilon_2(t))e^{i \omega t} a + h.c. ]$. 
The natural frequency is detuned by $\Delta=\omega_0-\omega$ from the driving. In the frame rotating at the driving frequency $\omega$, the Hamiltonian takes the form $H(t)= \Delta a^{\dagger} a  + \frac {\epsilon_1(t)} {2} (a+a^{\dagger})+ \frac {i \epsilon_2(t)} {2} (a-a^{\dagger})$. The dynamics in the presence of both dampings is accounted for by the Markovian Master equation~\cite{Lee13,Lee14}:
\begin{eqnarray}
	\dot{\rho} & = & - i [ H, \rho ]+ \kappa_1 \mathcal{D}[a^{\dagger}]\rho+ \kappa_2 \mathcal{D}[a^{2}]\rho 
	\label{eq:Masterequation}
\end{eqnarray}
with the Lindblad operator $\mathcal{D}[O]\rho=2 O \rho O^{\dagger} -  [ O O^{\dagger}, \rho ]_+$ where  $[,]_+$ refers to the anticommutator.  
Equivalently, we can recast the evolution in phase space with the Wigner distribution $W(\alpha,\alpha^*,t)$ equation in the coherent state representation~\cite{Lee13,CarmichaelBook}
\begin{eqnarray}
	\partial_t W & = & \{\mathcal{L}_H+ (\partial_{\alpha} \alpha+ \partial_{\alpha^{*}} \alpha^{*} ) [ -\kappa_1+2 \kappa_2 (|\alpha|^2-1 ) ] \nonumber \\
	& + &
	\partial_{\alpha}  \partial_{\alpha^{*}} [ \kappa_1+2 \kappa_2 (|\alpha|^2-1 ) ] \nonumber \\
		& + &  \frac {\kappa_2} {2} \left(	\partial^2_{\alpha} \partial_{\alpha^{*}} \alpha+ \partial_{\alpha} \partial^2_{\alpha^{*}} \alpha^*  \right) 	\} W
	\label{eq:Wignerequation}
\end{eqnarray}
where $\mathcal{L}_H = i \Delta ( \partial_{\alpha} \alpha - \partial_{\alpha^{*}} \alpha^{*} ) + \frac {i \epsilon_1} {2}  (\partial_{\alpha}-\partial_{\alpha^*}) + \frac {\epsilon_2} {2}  (\partial_{\alpha}+\partial_{\alpha^*})$ is the Liouvillian operator. Solving Eq.~\eqref{eq:Wignerequation} directly allows an independent verification of our results, and provides an interesting illustration of the wave-packet motion when the shortcut is applied (see below).

 From Eqs.~(\ref{eq:Masterequation},\ref{eq:Wignerequation}), the mean value $\langle \alpha \rangle_t =  {\rm Tr}[ \rho(t) a ]$ follows an equation analogous to Eq.~\eqref{eq:classicalvanderPol} written in the rotating frame
  \begin{eqnarray}
 \frac {d \langle \alpha \rangle} {d t}   & = &  - i \Delta  \langle \alpha \rangle +  (\kappa_1 +2 \kappa_2) \langle \alpha \rangle  -2 \kappa_2 \langle  |\alpha|^2 \alpha \rangle  \nonumber \\  &- &  \frac {1} {2} (\epsilon_2 +i\epsilon_1).
 \label{eq:average value equation}
 	\end{eqnarray}
The mean values $\langle \alpha^{m} \alpha^{* n} \rangle_t = \int d\alpha d\alpha^{*} \alpha^{m} \alpha^{* n} W(\alpha,\alpha^*,t) $ are taken with the Wigner distribution $W(\alpha,\alpha^*,t)$ in the coherent state representation, or equivalently $\langle \alpha^m \alpha^{* \: n} \rangle_t =  {\rm Tr}[ \rho(t) \mathcal{S} [ a^m a^{\dagger \: n} ] ]$ where $\mathcal{S}$ holds for the symmetric ordering of the creation/annihilation operators \cite{CarmichaelBook}. In the absence of driving and dissipative couplings, the system simply rotates at the frequency $\Delta$ under the influence of the detuning. Here, synchronization means that the driving $\epsilon$ is strong enough to prevent the system from being driven by the detuning $\Delta$. By solving Eq.~\eqref{eq:Masterequation}, the density matrix then converges to the steady solution centered on the position $\langle \alpha \rangle_{\infty} =  x_{\infty}+ i y_{\infty}$. The ratio between the dissipative couplings $\kappa_{1,2}$ dictates the average phonon number in the steady state, $\langle a^{\dagger} a \rangle_{\infty} = \kappa_1/2 \kappa_2+1$. This enables a clear distinction between the weakly ($\kappa_2 \ll \kappa_1$) and strongly ($\kappa_1 \ll \kappa_2$) nonlinear semi-classical regime. As in Ref.~\cite{Lee13},  we take the coupling values $(\kappa_1,\kappa_2)=(1,0.05)$ and $(\kappa_1,\kappa_2)=(0.05,1)$ when considering respectively the weakly and strongly nonlinear regimes. 

The difference between classical/quantum synchronization is also evident in the structure of their respective equations: while for classical dynamics Eq.~\eqref{eq:classicalvanderPol} is a closed differential equation, in the quantum case, Eq.~\eqref{eq:average value equation} couples the mean position $\langle \alpha \rangle$ to a hierarchy of moments
$\langle \alpha^m \alpha^{*n} \rangle$ by the presence of the term $\langle |\alpha|^2 \alpha \rangle$. Thus, in order to achieve perfect quantum synchronization, one must in principle match all these moments simultaneously to their steady -synchronized- values, which is generally intractable. Fortunately, as shown below, a strong acceleration can be obtained with shortcuts approaching simultaneously the target central position $\langle  x \rangle_{\infty},\langle  y \rangle_{\infty}$ and the third-order moment $\langle |\alpha|^2 x \rangle_{\infty},\langle |\alpha|^2 y \rangle_{\infty}$, that emerge at the lowest order in Eq.~(\ref{eq:average value equation}). Interestingly, the third-order moment may set a lower bound for the shortcut duration, and therefore constrain the quantum speed limit.

The initial density matrix corresponds to a coherent state, i.e. $\rho_0 = | \alpha_0 \rangle \langle \alpha_0 |$. As before, we design the drivings $\epsilon_i(t)$ on the time interval $0 \leq t \leq \tau$ (``shortcut'' part), and then we will fix their values $\epsilon_1(t)=1$ and $\epsilon_2(t)=0$. In the following, we choose the detuning $\Delta=2 \pi \times 0.05$, compatible with the onset of quantum synchronization. From Eq.~(\ref{eq:Masterequation}), we infer the stationary density matrix associated with the mean position $\langle \alpha \rangle_{\infty} =  x_{\infty}+ i y_{\infty}$, and third order moments  $\langle |\alpha|^2 x \rangle_{\infty},\langle |\alpha|^2 y \rangle_{\infty}$. Acceptable shortcuts should drive the quantum system to the mean position in the shortest possible time $\tau$, while providing third-order moments close to their stationary values. Inspired by the previous approach, we first build a shortcut using an inverse-engineering of the mean trajectory. To this end, we use an iterative procedure combined with a semi-classical approximation (truncation of the moments chain) $\langle  |\alpha|^2 \alpha \rangle \rightarrow |\langle  \alpha \rangle|^2 \langle  \alpha \rangle$ for the dynamical equation~\eqref{eq:average value equation}.  Under this assumption, the mean position $\langle \alpha \rangle$ follows Eq.~\eqref{eq:classicalvanderPol} with the substitutions:$\omega_0 \rightarrow \Delta$ and $\kappa_1 \rightarrow \kappa_1+2 \kappa_2$. Repeating the classical treatment, we set up a speeding up of the dynamics from the initial position $\alpha_0 = {\rm Tr} [ \rho_0 a]$  to the final position $  \langle \alpha \rangle_{\infty} $ in the required time interval $0 \leq t \leq \tau$. %($\rho_0$ is a 

\begin{figure}
  	\includegraphics[width=8.5 cm]{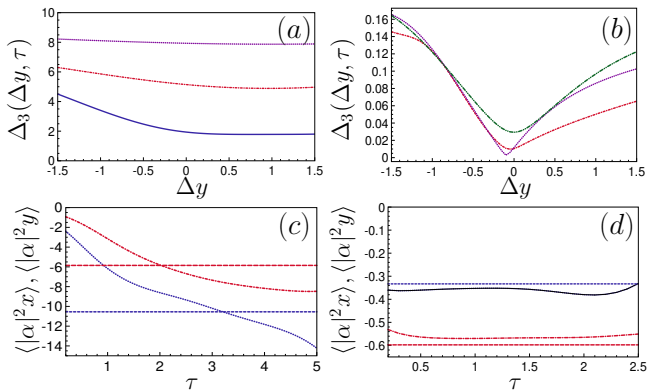}  
  	\caption{Distance of the 3rd-order moments to their target in a shortcut to quantum synchronization. (a,b): Mismatch $\Delta_3(\Delta y, \tau)$ [Eq.\eqref{eq:delta3}] of the final 3rd-order moments with respect to their targets as a function of the offset $\Delta y$ of the intermediate point $(x_m,y_m+\Delta y)$ for (a) the weakly and  (b) the strongly nonlinear regime. We have taken the shortcut durations $\tau=2$ (solid blue line, (a)), $\tau=1$ (red dash-dotted line, (a,b)),  $\tau=0.5$ (purple dotted line (a,b)) and $\tau=0.25$ (green dash-double-dotted line, (b)). (c,d): Third-order moments $\langle |\alpha|^2 x \rangle_{\tau}$ (red dash-dotted) and $\langle |\alpha|^2 y \rangle_{\tau}$ (blue dotted) as a function of the shortcut duration $\tau$ for the (c) weakly and (d) strongly nonlinear regimes. In (c,d) we have taken shortcuts corresponding to $\Delta y=0$. Horizontal lines represent their respective stationary values $\langle |\alpha|^2 x \rangle_{\infty}$ (red dashed) $\langle |\alpha|^2 y \rangle_{\infty}$ (blue dashed) under a constant drive $\epsilon_1=1,\epsilon_2=0.$ Details on the shortcut design can be found in Appendix B.}
  \label{fig:quantumsynchro}
\end{figure}

The following step consists in using a straight trajectory for the mean position  $\langle \alpha \rangle^{(1)}_t=( \langle \alpha \rangle_{\infty}  - \alpha_0) t /\tau$.  Such a solution introduced in Eq.~\eqref{eq:average value equation} under the semi-classical approximation, provides the driving functions $\epsilon_{1,2}^{(1)}(t)$. These functions are then used in the full quantum equation~\eqref{eq:Masterequation}. As a result, we find a final mean position slightly shifted from the target, i.e. $\langle \alpha \rangle_{\tau} = \alpha_{\infty} + \Delta \alpha^{(1)}.$ This offset can be corrected by iterating the procedure with a slightly modified target $\alpha^{(2)}_{\tau} =\langle \alpha \rangle_{\infty} - \Delta \alpha^{(1)}$ and a reference trajectory  $\alpha^{(2)}_t=( \alpha^{(2)}_{\tau}  - \alpha_0) t /\tau$.  After a few iterations, this approach leads to an improved shortcut trajectory for the full quantum problem with a driving of the mean position to a very close neighborhood of the target $\langle \alpha \rangle_{\infty}$ at the final time $\tau$. 
To bring also the third moments close to their target values $\{ \langle |\alpha|^2 x \rangle_{\infty},\langle |\alpha|^2 y \rangle_{\infty} \}$, we use the freedom in the choice of the trajectories connecting the initial/final points. In practice, we adjust both the shape and duration of the considered trajectory to get closer to these targets. Figures~\ref{fig:quantumsynchro}a,b show the influence of the trajectory shape on the final third-order moments for a few chosen shortcut durations in the weak and strongly nonlinear regimes.  Specifically, we use as reference trajectory two straight lines connecting the initial point $(x_0,y_0)$ to the target $(x_{\infty},y_{\infty})$ through an intermediate point $(x_m,y_m+\Delta y)$, with $(x_m,y_m)=(\frac 1 2 (x_0+x_{\infty}), \frac 1 2 (y_0+y_{\infty}))$ are the middle-point coordinates and $\Delta y$ is an offset. Each segment is followed at a constant speed for half the total shortcut time. To evaluate the effectiveness of our protocol, we introduce the distance 
 \begin{equation}
	\label{eq:delta3}
	\Delta_3(\Delta y, \tau) = \left[ \sum_{j=x,y} \left(\langle |\alpha|^2 j \rangle_{\tau}-\langle |\alpha|^2 j \rangle_{\infty} \right)^2 \right]^{1/2}
	\end{equation}
that depends on the chosen shortcut path and duration.  The driving amplitudes $\epsilon_{1,2}(t)$ are obtained from the previous procedure based on Eq.~(\ref{eq:average value equation}) and a semi-classical approximation~(See Appendix B for details).  In Figs.~\ref{fig:quantumsynchro}c,d, we plot the third-order moments 	$\langle |\alpha|^2 x \rangle_{\tau}, \langle |\alpha|^2 y \rangle_{\tau}$ and their stationary values as a function of the total duration $\tau$ for a given trajectory. 
 
In the weak linear regime, the final third-order moments are much more sensitive to the shortcut duration than to the trajectory shape. Indeed, the system behaves in this case as a driven harmonic oscillator, for which the shape of the trajectory has no direct influence on the final wave-packet width. A minimum expansion time is required to approach the correct third-order moments, which sets a lower bound on acceptable shortcut durations, and thus constrain the quantum speed limit in this regime. 

In contrast, in the strongly nonlinear regime, the path shape has a more drastic influence than the shortcut duration. Moreover, the final third-order moments can be approached in a time $\tau=0.25$ that is an order of magnitude faster than the time scale $\tau \simeq 2$ of the weakly nonlinear regime. This faster convergence originates from the fact that the distribution of the stationary density matrix has a width closer to that of the initial coherent state ($\Delta \alpha = 0.5$), as the wave function experiences a sharper confinement to the circle of radius $|\alpha_{\infty}|$. In general, for a given trajectory, there is no $\tau$ duration that gives a perfect simultaneous match of the two third-order moments $\langle |\alpha|^2 x \rangle_{\tau},\langle |\alpha|^2 y \rangle_{\tau}$ with their respective targets. However, this can occur in specific trajectories, such as the one associated to $\Delta y=-0.1$ with the duration $\tau \simeq 0.5$. These  specific trajectories are excellent candidates to build an efficient shortcut to quantum synchronization. Nevertheless, even when such trajectories are unavailable, a small mismatch in the third-order moments is actually not critical for the success of the protocol. The system dynamics turns out to be mainly driven by the fifth (and higher)-order moments once the mean position and third-order moments are below a certain distance from their targets.

 %we discuss below examples where these moments are slightly shifted from their target values, but where the shortcut still provides a significant acceleration to quantum synchronization
  
\begin{figure}[htbp]
			\includegraphics[width= 8 cm]{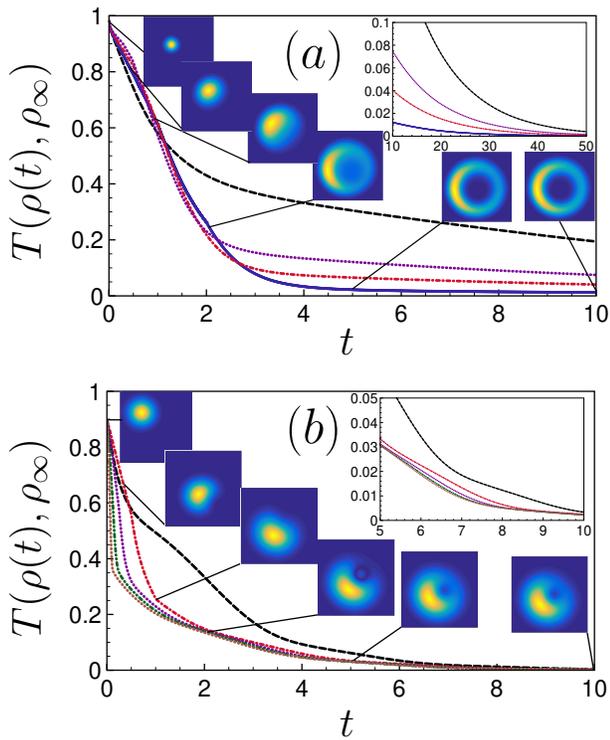}
	\caption{Performance of the shortcut to quantum synchronization: 
		(a,b): Trace distance $T(\rho(t),\rho_{\infty})$ of the density matrix $\rho(t)$ to the stationary solution $\rho_{\infty}$ as a function of time for the weak (a) and strong (b) nonlinear regimes with the respective shortcuts of Fig.2a and Fig.2b with $\Delta y=0$.  We have taken the durations $\tau=2$ (solid blue line, (a)),  $\tau=1$ (red dash-dotted line, (a,b)), $\tau=0.5$ (dotted purple line, (a,b))  $\tau=0.25$ (green dash-double dotted line, (b)) and $\tau=0.125$ (brown dotted line, (b)). The black-dashed line stands for a constant drive of amplitude $\epsilon_1=1$ and $\epsilon_2=0$. The insets represent the Wigner distribution $|W(x,y,\tau)|$ ($x={\rm Re}[\alpha]$ and $y={\rm Im}[\alpha]$) for (a) the shortcut designed with $\tau=2$ in the weakly nonlinear regime and for (b) the shortcut designed with $\tau=0.5$ in the strongly nonlinear regime for the considered times $t=0,0.5,1,2,5,10$.}
	\label{fig:quantumperformance}
\end{figure}

Ideally, the density matrix after the shortcut to quantum synchronization should coincide with the stationary solution. To estimate quantitatively the shortcut performance, we use the trace distance~\cite{NielsenChuangBook} between the instantaneous density matrix of the system and the stationary solution for a sudden and constant driving. This metric is defined as $T(\rho_1,\rho_2)= \frac {1} {2} \sum_{i=1}^{N} |\lambda_i|$ with $\{\lambda_1,..., \lambda_N \}$ the eigenvalues of the matrix $\rho_1-\rho_2$.
   A faster decay of the trace distance is indicative of an accelerated quantum synchronization.  Figures~\ref{fig:quantumperformance}a,b compare the evolution of the trace distance for a shortcut protocol and for a sudden and constant driving. These figures show that for a constant driving, synchronization occurs at a faster rate in the strongly nonlinear regime where quantum noise has a greater influence. This observation corroborates the role played by quantum noise in the building up and acceleration of quantum synchronization discussed in Ref.~\cite{QuantumSynchro22noise}. Figures \ref{fig:quantumperformance} reveal that in both weakly and strongly nonlinear regimes, the shortcut accelerates the decay of the trace distance.  However, significant differences are observed between both regimes regarding the overall shortcut performance.  First, a more drastic speed-up is obtained from the shortcut in the weakly nonlinear regime. Moreover, the resulting acceleration is then noticeably larger for the shortcut of duration $\tau=2$ - while the shortcut performance is significantly reduced when smaller durations ($\tau=0.5$, $\tau=1$) are used. Indeed, the trace distance falls below the $1\%$ threshold at time $t \simeq 11.9$ for the shortcut of duration $\tau=2$, against $t=24.6$, $t=31.0$, $t=40.4$ respectively for the shorcuts of durations ($\tau=1$, $\tau=0.5$) and for the constant drive. In comparison, for the strongly non-linear regime, the shortcut duration has almost no influence on the decay of the trace distance - showing that all shortcuts have similar performance regardless of the chosen duration. This finding is consistent with the third-order moments sketched in Fig.~\ref{fig:quantumsynchro}(a,b): in the weakly non-linear regime the shortcut $\tau=2$ is the only one that yields final third-order moments in a neighborhood of their target, while in the strongly non-linear regime the final third-order moments is barely affected by the shortcut duration. This strongly suggests that a proper matching of the third-order moments is necessary to accelerate quantum synchronization. In the quasi-linear regime, this sets a lower bound on quantum synchronization time. In contrast, in the strongly non-linear regime, the shortcut durations can be chosen arbitrarily small. Naturally, the resources employed (amplitude of the time-dependent drives $\epsilon_1(t),\epsilon_2(t)$) increase as the duration $\tau$ is reduced, and the maximum accessible driving amplitude will eventually impose a minimum duration $\tau$.  These considerations can be connected to a more general discussion on quantum speed limits in both linear and nonlinear quantum systems: they suggest that the presence of nonlinearities can provide acceleration of quantum protocols by acting on the shape of propagating quantum wave-packets.
 
 Finally, we illustrate our results by sketching the Wigner distribution at different stages of the shortcut protocol as insets of Figures~\ref{fig:quantumperformance}a,b. These pictures provide qualitative insights on the convergence towards the quantum-synchronized Wigner distribution. In both the weakly/nonlinear regimes, the initial distribution is Gaussian-shaped centered on $\alpha_0=-1+i$.  The insets show that the Wigner distributions delivered by the shortcut share several features with the target density matrix (insets of Figs.~3a,b, $t=10$) - such as a squeezing in the amplitude $|\alpha|,$  and a phase locking  corresponding to a preferred phase $\phi$ ($\alpha=|\alpha| e^{i \phi}$) associated to the center of the Wigner distribution. In the weakly non-linear regime, the Wigner distribution profile evolves rapidly towards a ring shape.  The strong similarity between the Wigner distribution immediately after the shortcut (inset of Fig.3a, $t=2$) and the target explains the rapid convergence witnessed by the trace distance. In the strongly non-linear regime, similar features between the instantaneous distribution and the target -- such as the presence of a ``hole'' associated to a low-probability zone -- also appear progressivly after termination of the shortcut. The Wigner distributions were obtained from a numerical resolution of  Eq.~(\ref{eq:Wignerequation}) based on a Crank-Nicholson scheme on a $2^8 \times 2^8$ grid with a time step $\delta t=5 . 10^{-4}$~\cite{PRL20}. Both numerical methods, the partial differential equation~\eqref{eq:Wignerequation} and the Master equation~\eqref{eq:Masterequation}, agrees for the mean positions $\langle x \rangle_t,\langle y \rangle_t$ across the whole considered time interval with an accuracy better than $0.1$ \%. The Wigner phase-space simulation thus provides an additional independent confirmation of the effectiveness of the proposed shortcut drivings.
  
  % The insets in Figs.~\ref{fig:quantumperformance} reveal that the Wigner distributions delivered by the shortcuts share several qualitative features with the target density matrix   associated to the quantum-synchronized state - such as the squeezing in the amplitude $|\alpha|,$ or the presence of a phase-locking corresponding to a preferred phase $\phi$ ($\alpha=|\alpha| e^{i \phi}$) associated to the center of the Wigner distribution.  \DGO{DISCUSSION A REPRENDRE AVEC LE DESSIN DE LA WIGNER INITIALE SUR LES FIGURES}The similarity between the Wigner distributions of Fig.4a and  Fig.4b, which expose the strong acceleration in the trace distance decay (Fig.3c, blue line). The Wigner distributions of Fig.4c and Fig.4d have a more different shape, which explains why the acceleration provided by the shortcut is less impressive in this example (Fig.3d, red dashed-dotted line).  These qualitative elements are an additional indication of the fast convergence towards quantum synchronization.

 To summarize, we have detailed a systematic procedure, inspired by shortcut-to-adiabaticity techniques, allowing synchronization acceleration in both classical and quantum systems. Classical synchronization has been discussed in the context of the Van der Pol oscillator. By using an appropriate time-dependent driving amplitude instead of the usual constant profile, one can accelerate the oscillator motion from a given initial point to the limit cycle. In a driven quantum Van der Pol oscillator, reaching an exact quantum synchronization requires to make coincide the system density matrix and the stationary solution associated to quantum synchronization. We have developed a shortcut strategy to ensure a ``quasi''-synchronized state based on the simultaneous control of the mean position and of the third-order moments of the quantum oscillator. Our results show a different behavior in the weakly and strongly nonlinear regimes: in the latter, the third-order moments depend on the trajectory shape, and shortcuts of faster durations can be employed to reach the approximately synchronized state when compared to the quasi-linear regime. The method presented here could be adapted to other non-linear quantum systems for which the control of the wave-function shape is critical.  

  \acknowledgments{This study has been supported through the EUR grant NanoX No. ANR-17-EURE-0009 in the framework of the ``Programme d'Investissements d'Avenir" and the
Agence Nationale de la Recherche through grant ANR-18-CE30-0013. F.I. acknowledges funding from the Brazilian agencies Conselho Nacional 
de Desenvolvimento Científico e Tecnológico (310265/2020-7), Coordena\c{c}\~ao de Aperfei\c{c}oamento de 
Pessoal de N\'ivel Superior (Programa CAPES-PRINT), and Funda\c{c}\~ao de Amparo a Pesquisa do Estado do Rio de Janeiro~(210.296/2019). This work is part of the INCT-IQ (465469/2014-0).}

%\FI{The authors are grateful to Romain Duboscq for a previous collaboration on guided atomic waves}.

\appendix

\section*{APPENDIX A: SHORTCUT-TO-SYNCHRONIZATION IN A CLASSICAL VAN-DER-POL OSCILLATOR}

We detail here the procedure to design of a shortcut-to-synchronization from the initial point $(x_0,y_0)=(0,0)$. We first solve the trajectory of a sinusoidally driven van-der-Pol oscillator with $\epsilon_0=1.5/T_0$ and obtain the branching point $(x_{\infty},y_{\infty}) \simeq (0.29,1.05)$ corresponding to $t_{\infty}=50.125 \times T_0$. The proximity of this branching point to the ``vertical'' of the initial position $(x_0,y_0)$ enables a fast shortcut with the amplitude $\epsilon(t)$.
 %We verify below that this point can be connected to $(x_0,y_0)$ via a shortcut of the given duration $\tau=T/4$. This choice of $t_{\infty}$ is arbitrary and by no means unique. All conclusions below are, by construction, independent of the time-scale $T$.

We first define a system trajectory of the form $y_{{\rm short}, \gamma}(t)=  P_{\gamma}(t/\tau)$ with the polynomial $P_{\gamma}(u)= y_{\infty}+ y_0'(t_{\infty}) \tau (u-1)+(y_0-y_{\infty}+ y_0'(t_{\infty}) \tau )(u-1)^2+\gamma u(u-1)^2$. The chosen trajectory fulfills, for any value of the parameter $\gamma$, the required boundary conditions $y_{{\rm short},\gamma}(0)=y_0$ and $y_{{\rm short}, \gamma}(\tau)=y_{\infty}$ associated respectively to the initial and final shortcut times. The additional condition $y_{{\rm short},\gamma}'(\tau)=y_0'(t_{\infty})$ provides a continuity of the driving amplitude between the transient and sinusoidal part.

To determine the correct shortcut trajectory and fix the $\gamma$ parameter, we use a self-consistency argument: by virtue of Eq.\eqref{eq:classicalvanderPol}, when the system goes along the trajectory $y_{{\rm short},\gamma}(t)$, the coordinate motion $x(t)$  follows a differential equation where  $y_{{\rm short},\gamma}(t)$ acts as a driving term:
\begin{equation}
	\dot{x}= \omega_0 y_{{\rm short}, \gamma} + \kappa_1 x - 2 \kappa_2 (x^2+ y_{{\rm short}, \gamma}^2)x.
\end{equation}
With the considered initial condition $x(0)=x_0=0$, each value of $\gamma$ yields a corresponding solution $x_{{\rm short},\gamma}(t)$ and final coordinate $x_{{\rm short},\gamma}(\tau)$ at the time $\tau$. For the ``magic''value $\gamma_0$, the final coordinate reaches the target, i.e.  $x_{{\rm short},\gamma_0}(\tau)=x_{\infty}.$ Then, the trajectory $(x_{{\rm short},\gamma_0}(t),y_{{\rm short},\gamma_0}(t))$ reaches the branching point $(x_{\infty},y_{\infty})$ at $t=\tau$, and can thus be choses as shortcut trajectory. For the parameters above, one finds numerically $\gamma_0 \simeq -9.3532$. The corresponding driving amplitude $\epsilon_{\rm short}(t)$ is derived from Eq.~\eqref{eq:classicalvanderPol} as
\begin{eqnarray}
	\label{eq:driving_amplitude_classical}
\epsilon_{\rm short}(t)  &= & -2[\dot{y}_{\rm short, \gamma_0}(t)+\omega_0 x_{\rm short, \gamma_0}(t)-\kappa_1 y_{\rm short, \gamma_0}(t) \nonumber \\
	  &+ &  2 \kappa_2 (x_{\rm short, \gamma_0}^2(t) + y_{\rm short, \gamma_0}^2(t)) y_{\rm short, \gamma_0}(t) \nonumber \\
	  & + & y_{\rm short, \gamma_0}(t) ] 
\end{eqnarray}
for $t \leq \tau$. For $t > \tau$, the sinusoidal driving is resumed $\epsilon(t) = \epsilon_0 \cos( \omega t + \varphi)$. The  phase $\varphi$ is fixed as follows. At time $\tau$, the system is at a position that would be reached under a plain sinusoidal driving $\epsilon_0(t)$ at time $t_{\infty}$. For our strategy to be sucessful, the system must be subject to a driving $\epsilon(t)$ such that $\epsilon(t-\tau)=\epsilon_0(t-t_{\infty})$ for $t > \tau$. A suitable choice is thus $\varphi = \omega (t_{\infty}-\tau).$\\

\section*{APPENDIX B: SHORTCUT-TO-SYNCHRONIZATION IN A QUANTUM VAN-DER-POL OSCILLATOR}

We detail the procedure for the shortcuts considered in Figs.~2 and 3 in the weakly/strongly non-linear regimes. We solve Eq.~\eqref{eq:Masterequation} in a quantum subspace corresponding to the $N$ lowest-energy level of the harmonic oscillator. It is sufficient to consider $N=40$, as higher-energy quantum states are irrelevant for the considered initial states and Hamiltonians.

In the weakly nonlinear regime, for a sinusoidal driving with $\epsilon_1=1$, $\epsilon_2=0$, one finds the stationary mean position $\alpha_{\infty}=x_{\infty}+i y_{\infty} \simeq -0.86 -0.38 i$ and the corresponding middle-point $\alpha_{m}= \frac 1 2 (\alpha_0+\alpha_{\infty}).$ For a generic intermediate point $\alpha_I= \alpha_m+ i \Delta y,$ we use a piece-wise defined path $\langle \alpha \rangle_t^{(1)}= \alpha_0+2 (\alpha_I-\alpha_0) t / \tau$ for $0 \leq t \leq  \tau / 2$ and $\langle \alpha \rangle_t^{(1)}= \alpha_I +2 (\alpha_{\infty}-\alpha_I) t / \tau$ for $\tau / 2 \leq t \leq \tau$. The driving amplitudes $\epsilon_{1,2}^{(1)}(t)$ can be expressed from Eq.~\eqref{eq:average value equation} with the semiclassical approximation:
\begin{eqnarray}
& &  \frac 1 2 \left( \epsilon_2^{(1)}(t)  +  i \epsilon_1^{(1)}(t) \right)    =   	- \frac {d \langle \alpha \rangle_t^{(1)}} {d t}  - i \Delta  \langle \alpha \rangle_t^{(1)}  \nonumber\\
 & \: & \qquad   + (\kappa_1 +2 \kappa_2) \langle \alpha \rangle_t^{(1)}    -  2 \kappa_2 |\langle  \alpha \rangle_t^{(1)}|^2  \langle  \alpha \rangle_t^{(1)} 
	\label{eq:average value equation2}
\end{eqnarray}
As an example, we consider the weakly nonlinear regime with a shortcut duration $\tau=2$ and $\Delta y=0$. The quantum trajectory on the time interval $0 \leq t \leq \tau / 2$ starts at the initial point $\alpha_0=-1+i$ and ends at the intermediate point $\alpha_m\simeq -0.93+0.34 i$. A numerical resolution of Eq.\eqref{eq:Masterequation} with the amplitudes $\epsilon_{1,2}^{(1)}(t)$ yields a first offset $\Delta \alpha^{(1)}= \langle x \rangle_{\tau/2} - x_m + i ( \langle y \rangle_{\tau/2} - y_m)$. As indicated in the main text, we iterate the procedure with corrected trajectories $\langle \alpha \rangle_t^{(n)}= \langle \alpha \rangle_t^{(n-1)} - \Delta \alpha^{(n-1)}$. From successive iterations, one obtains $\Delta \alpha^{(1)} \simeq 0.34-0.27 i,$ $\Delta \alpha^{(2)} \simeq 0.038 -0.041 i $ and  $\Delta \alpha^{(3)} \simeq (2.2-5.2 i) \times 10^{-3}$. For the strongly non-linear regime with a shortcut of duration $\tau=0.5$ and $\Delta y=0$, considering the intermediate point $\alpha_m \simeq -0.63+0.31 i$ and time interval  $0 \leq t \leq \tau / 2$, our procedure delivers the successive offsets  $\Delta \alpha^{(1)} \simeq 0.23  - 0.18 i $, $\Delta \alpha^{(2)} \simeq  (-8.6+8.0  i)\times 10^{-3} $, $\Delta \alpha^{(3)} \simeq (6.1 - 3.4 i) \times 10^{-4} $. The convergence is fast and a few iterations are sufficient for the purpose of driving the mean position close to its target. In our example, after three iterations the error on the mean position becomes irrelevant: the speed of quantum synchronization is then mostly limited by a mismatch in the third (and higher-order) moments with respect to their stationary values. The convergence of the iterative process increases when shorter time intervals are considered between the initial and intermediate points. For a given duration, the convergence is faster in the weakly nonlinear regime - the semi-classical approximation is more accurate in this case.

\end{document}